\documentclass[pra,twocolumn]{revtex4}
\usepackage{amsmath}
\usepackage{amsfonts}
\usepackage{graphicx}

\begin{document}

\title{Dynamics of periodic anticrossings: Decoherence, pointer states and hysteresis curves}
\author{P\'{e}ter F\"{o}ldi}
\affiliation{Department of Theoretical Physics, University of Szeged, Tisza Lajos k\"{o}r%
\'{u}t 84, H-6720 Szeged, Hungary}
\author{Mih\'{a}ly G. Benedict}
\affiliation{Department of Theoretical Physics, University of Szeged, Tisza Lajos k\"{o}r%
\'{u}t 84, H-6720 Szeged, Hungary}
\author{F. M. Peeters}
\affiliation{Departement Fysica, Universiteit Antwerpen, Groenenborgerlaan 171, B-2020 Antwerpen, Belgium}

\begin{abstract}
We consider a strongly driven two-level (spin) system, with a periodic external field that induces a sequence of avoided level crossings. The spin system interacts with a bosonic reservoir which leads to decoherence. A Markovian dynamical equation is introduced without relying on the rotating wave approximation in the system-external field interaction. We show that the time evolution of the two-level system is directed towards an incoherent sum of periodic Floquet states regardless of the initial state and even the type of the coupling to the environment. Analyzing the time scale of approaching these time-dependent pointer states, information can be deduced concerning the nature and strength of the system-environment coupling. The inversion as a function of the external field is usually multi-valued, and the form of these hysteresis curves is qualitatively different for low and high temperatures. For moderate temperatures we found that the series of Landau-Zener-St\"{u}ckelberg-type transitions still can be used for state preparation, regardless of the decoherence rate. Possible applications include quantum information processing and molecular nanomagnets.
\end{abstract}

\maketitle

\section{Introduction}
Crossings and anticrossings of energy levels play important role in various physical systems. The case when the level scheme is time dependent is of special interest, because the separation of the levels strongly influences the dynamics, e.g., depending on the parameters, transitions may occur around an anticrossing. For a two-level system with a linearly time-dependent Hamiltonian, the analytically solvable Landau-Zener-St\"{u}ckelberg (LZS) model \cite{L32,Z32,S32} reflects the most important properties of the dynamics. Phenomena that can be described using this model include the dynamics of molecular vibrations \cite{MK01,LS93}, multiphoton transitions and ionization of Rydberg atoms \cite{LG84,PG84,SDG94,GWG95}, molecular nanomagnets \cite{WS00,CB03,WB05,GSV06} and also quantum information processing with superconducting qubits \cite{SH06,BO06}. Multilevel and nonlinear generalizations of the LZS model have also been studied extensively (see e.g.~\cite{VS99,GS02,PP07,VIV07}). The LZS model itself can describe the adiabatic limit, when the system follows the instantaneous eigenstates of the Hamiltonian, as well as the case when a sudden transition takes place. Generally (between the two extreme situations mentioned above), an LZS transition splits the state of the system into two parts, which are almost orthogonal when the transition region is left. In this sense, considering periodic driving  that forces the system to return to the transition region after a half-cycle, fundamental which-way interference effects \cite{LL65} can appear. This kind of interference is highly sensitive to dephasing and other decoherence mechanisms \cite{GJK96}, which are the phenomena to be investigated in the present paper.

We consider a two-level system driven by an external field leading to periodic anticrossings. Mathematically, this is similar to the LZS model with harmonic terms replacing the usual linear ones in the Hamiltonian. The spin $1/2$ system is assumed to be embedded in a thermal reservoir, and using Floquet theory \cite{F883} we introduce the appropriate master equation in a systematic way. Investigating the dynamics induced by the master equation we determine the direction and time scale of the decoherence. We find that the time-dependent Floquet states play the role of pointer states \cite{Z81}. The temperature dependence of the process is also investigated with a special focus being on the population difference (that is, $\langle \sigma_z \rangle$) as a function of the external field. In the context of molecular nanomagnets, similar magnetization curves have recently been investigated in pulsed fields both experimentally (see e.g.~\cite{CB00,CB03,RL05b}) and theoretically \cite{RL05,VG06}. Our findings in the periodic case generalize the result that the multi-valuedness of the hysteresis curves in these systems reflects the relaxation towards the quasistationary (equilibrium) solutions, which are related to single-valued magnetization curves.

\section{Dynamical equations}
Dynamics of strongly driven open systems requires special attention, as the external field modifies not only the Hamiltonian of the investigated system, but also its interaction with the environment. Here we consider a periodically driven two-level system that exhibits a series of avoided level crossings. This "periodic version" of the Landau-Zener-St\"{u}ckelberg model can be described by a Hamiltonian
\begin{equation}
H_{s}(t)=a \cos(\Omega t)\sigma_z +\frac{\delta}{2}\sigma_x=\left(
\begin{array}{cc}
a \cos(\Omega t) & \delta/2 \\
\delta/2 & -a \cos(\Omega t)
\end{array}
\right),  \label{Hs}
\end{equation}
where $\Omega=2\pi/T$ is the angular frequency of the external field and the second equality holds in the $\{|+\rangle,|-\rangle\}$ eigenbasis of $\sigma_z$.
Although similar model Hamiltonians can describe a large variety of physical systems, for the sake of definiteness we will consider a spin $1/2$ system being coupled via Zeeman interaction to an oscillating magnetic field. We assume that the spin is embedded in a bosonic reservoir (which, in this aspect, can represent phonons if the distinguished spin is coupled to a crystal lattice) described by the Hamiltonain
\begin{equation}
H_{r}=\sum_k \omega_k {a_k}^{\dagger} a_k,  \label{Hr}
\end{equation}
where $a_k$ and ${a_k}^{\dagger}$ are the annihilation and creation operators of the $k$-th mode satisfying $\left [ a_{k^{\prime}}, {a_k}^{\dagger}\right ]=\delta_{kk^{\prime}}.$ (Note that we set $\hbar=1.$)
The system-environment interaction Hamiltonian is written as
\begin{equation}
V=\mathcal{S} \otimes \sum_k g_k ({a_k}^{\dagger} + a_k),  \label{V}
\end{equation}
where $\mathcal{S}$ can represent any (hermitian) spin operator, and the coupling constants $g_k$ are assumed to be real. (Note that this kind of interaction resembles the coupling of a two-level atom to electromagnetic field modes \cite{WM94}, as well as to spin-phonon coupling in solids \cite{LL00}.)

Considering only the periodic $H_{s}$ as the generator of the time evolution (no environmental influence), Floquet theory tells us that it is possible to find a time-dependent eigenbasis
\begin{eqnarray}
&|\phi_r (t)\rangle=|u_r(t)\rangle e^{-i \epsilon_r t}, \ \ |u_r(t+T)\rangle=|u_r(t)\rangle,
\nonumber \\ &\langle  u_1(t)|u_2(t)\rangle=0,  \ \ \langle u_r(t)|u_r(t)\rangle=1. \label{basis}
\end{eqnarray}
Unlike the Floquet states $|u_r(t)\rangle,$ the elements of this basis themselves are are not $T$-periodic functions, which is related to a nontrivial phase effect \cite{WS99}. Let us recall that if $\epsilon_r$ is a Floquet quasi-energy and the corresponding state is $|\phi_r (t)\rangle,$ then the same holds for $\epsilon_r + n\Omega$ and $|\phi_r (t)\rangle\times\exp(i n \Omega t)$ for any integer $n.$ However, these states are equivalent from the dynamical point of view, thus it is sufficient to focus on the two nonequivalent quasi-energies the magnitude of which are the closest to zero. For the sake of definiteness we will assume $\epsilon_1\leq \epsilon_2.$ Note that the Schr\"{o}dinger equation induced by the Hamiltonian (\ref{Hs}) can be rewritten as an inhomogeneous differential equation of Mathieu-type, numerical methods for computing $\epsilon_1$ and $\epsilon_2$ using this fact can be found e.g.~in Ref.~\cite{AS65}.  Having obtained the quasi-energies and the corresponding states given by Eq.~(\ref{basis}), the time evolution operator $U(t)= \sum_r |\phi_r (t)\rangle \langle \phi_r (0)|$ can be constructed. Then, returning to the open system, we can transform $V$ into an interaction picture with $H_r$ and $H_s$ being the Hamiltonians of the uncoupled total system. In this way, we obtain:
\begin{equation}
\mathcal{S}_I(t)=\sum_{\omega>0}S(\omega) e^{-i \omega t} + h.c., \label{Vt}\end{equation}
where
\begin{eqnarray}
S(\omega)= \sum_{r,r^\prime,n}|u_r(0)\rangle \langle u_{r^\prime}(0)| \langle \langle r'|S|r\rangle\rangle_n , \\
 \langle \langle r'|S|r\rangle\rangle_n=\frac{1}{T}\int_{0}^{T}  e^{i n \Omega t}\langle u_{r^\prime}(t)|\mathcal{S}|u_r(t)\rangle d t, \label{Vomega}
\end{eqnarray}
with $n$ being integer, and the sum runs over indices satisfying $\epsilon_{r^\prime}-\epsilon_r-\Omega n =\omega$ \cite{BP02}.  This means that -- unless the Fourier component $S(\omega)$ is zero -- three sets of positive frequencies appear,
\begin{equation}
\omega^0_{n}= n \Omega , \ \ \omega^\pm_n= \pm |\epsilon_1-\epsilon_2| + n \Omega .
\end{equation}
Assuming that initially the density operator of the complete system factorizes, $\rho(0)=\rho_s(0)\otimes \rho_r(0)$, standard methods (Born-Markov approximation) lead to the interaction picture master equation
\begin{equation}
\frac{d}{d t}\rho_s(t)=-\mathrm{Tr}_r{\int_0}^T\left[V(t),\left[V(t-s),\rho_s(t)\otimes \rho_r\right]\right] d s,
\label{intmast}
\end{equation}
where $\mathrm{Tr}_r$ means trace over the reservoir degrees of freedom, and the explicit indication of the interaction picture has been omitted. Next we insert Eq.~(\ref{Vt}) with the standard periodic ($\exp(\pm i \omega_k t)$) time dependence of the interaction picture creation and annihilation operators of the bath into the master equation above. Then the sums over environmental modes appearing in Eqs.~(\ref{V}) and (\ref{Hr}) are transformed into an integral over the mode frequencies with the mode density $\mathcal{D}(\omega)$ being a weight factor. Finally we perform rotating wave approximation (RWA) in the system-environment interaction \cite{BP02}. (Note that we did not use RWA for the coupling of the system and the external field.) With this approximation the integral in Eq.~(\ref{intmast}) can be evaluated, leading to an interaction picture Born-Markov master equation for the reduced density operator of the two-level system \cite{HP97}:
\begin{eqnarray}
& &\frac{d \rho_s}{d t}= \sum_{\omega>0} \gamma(\omega) \nonumber \\ & &\times \left( S^{\dagger}(\omega) \rho_s S(\omega) -\frac{1}{2} S(\omega)S^{\dagger}(\omega) \rho_s
-\frac{1}{2} \rho_s S(\omega)S^{\dagger}(\omega) \nonumber \right) \\
& &+ \gamma'(\omega) \left( S(\omega) \rho_s S^{\dagger}(\omega)  -\frac{1}{2} S^{\dagger}(\omega)S(\omega) \rho_s
\nonumber
\right.\\ & & \left.-\frac{1}{2} \rho_s S^{\dagger}(\omega)S(\omega) \right),\label{master}
\end{eqnarray}
where
\begin{eqnarray}
\gamma(\omega)&=&\mathcal{D}(\omega) g^2(\omega) \left(\langle n(\omega) \rangle +1\right), \nonumber\\ \gamma'(\omega)&=&\mathcal{D}(\omega) g^2(\omega) \langle n(\omega) \rangle  \label{gammas},
\end{eqnarray}
with $\langle n(\omega) \rangle=\mathrm{Tr}_r \left(a^\dagger(\omega) a(\omega) \rho_r\right)$ representing the average number of excitations in the environmental mode labeled by the frequency $\omega.$ Returning to the Schr\"{o}dinger picture, Eq.~(\ref{master}) turns out to be a Lindblad-type equation \cite{L76}, but as a consequence of the strong driving field, the Lindblad operators have explicit time dependence. We note that decoherence effects can also be described by the aid of the Feynman-Vernon influence functional method \cite{FV63}, which has already been applied successfully \cite{TJ97,TF98} to bistable quantum systems (like the one considered here).

Additionally, as a result of the system-environment RWA, the frequencies appearing in Eqs.~(\ref{Vt}) and (\ref{master}) are exactly the same. Recalling that if $\epsilon_r$ is a Floquet quasi-energy, then the same holds for $\epsilon_r + n\Omega$ for any integer $n$, it is possible to establish a connection between the Floquet spectrum  and the combined energy levels of a two-level system and a quantized single mode field \cite{BP02}. In view of this, the master equation above can be interpreted as the coupling of those transitions of the combined spin-field systems to the resonant reservoir mode, where the matrix element of the coupling operator is nonzero. Technically, it is convenient to collect the terms in Eq.~(\ref{master}) that contain the same operator part (eg. $|u_1 (0)\rangle \langle u_2(0)|$) then to calculate their common coefficient by evaluating the sum over frequencies for a sufficiently large, but practically finite number of modes (see subsec.~\ref{couplings}). Note that interaction with the environment usually also induces a Lamb-type renormalization of the energy spectrum of the system, but in this framework the renormalization Hamiltonian (that should appear on the rhs. of Eq.~(\ref{master}) in a commutator with $\rho_s$) commutes with the system Hamiltonian $H_s$ \cite{BP02}, thus it has no special importance from our point of view.

Finally, let us note that in the optical case (when the offdiagonal elements of $H_s$ oscillate) with system-field RWA and exact resonance the Floquet quasi-energies and states can be calculated analytically, see Refs.~\cite{HP97,BP02}, where the quantum jumps leading to the strong-driving Mollow spectrum are also introduced.

\section{Direction and characteristic time of decoherence}
\subsection{Summary of the free time evolution}
Before investigating how the environment induced decoherence modifies the dynamics of the spin system, it is certainly worth recalling the main features of its free (unitary) time evolution. First, let us note that by exchanging $\sigma_x$ and $\sigma_z$ in the Hamiltonian (\ref{Hs}), we obtain a hermitian operator $\widetilde{H_s}$ that is unitarily equivalent to the original system Hamiltonian: $H_s=1/2(\sigma_z+\sigma_x)\widetilde{H_s}(\sigma_z+\sigma_x)$. As $\widetilde{H_s}$ can describe a two-level atom subjected to electromagnetic field (classical Rabi problem without RWA \cite{WM94}), several studies have been devoted to the time evolution induced by this Hamiltonian. Floquet analysis of the dynamics has already been done in Ref.~\cite{S65}, while more recent results (see \cite{SH06,BO06,AN07} and references therein) usually focus on the applications in quantum information processing.
Considering a different field where the model above appears, let us note that as an additional effect underlying the analogy between atoms and quantum dots, exciton Rabi oscillations have also been detected \cite{SXS01,UW05}. In these strongly confined solid state systems, however, memory effects are important, the dynamics can often be non-Markovian \cite{VP07}.

%%%%%%%%%%%%%%%%%%%%%%%%%%%%%%%%%%%%%%%%%%%%%%%%%%%%%%%%%%%%%%%%%%%%%%%%%%%%
\begin{figure}[tbp]
\begin{center}
\includegraphics*[width=8.5 cm]{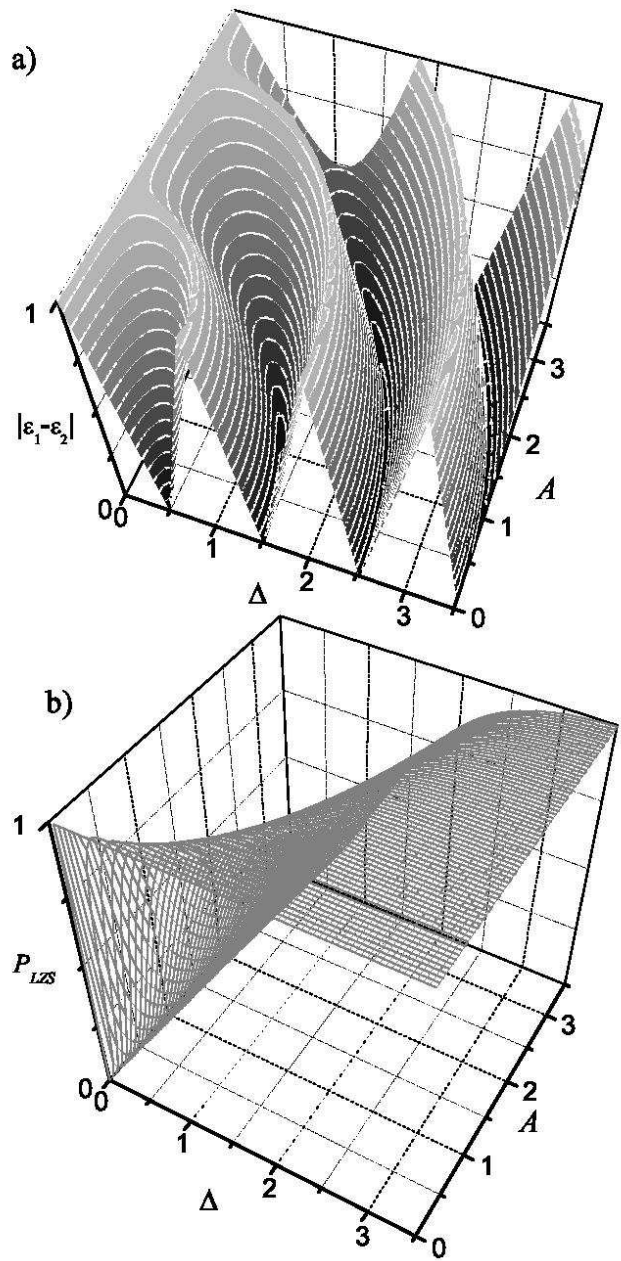}
\end{center}
\caption{a) The difference of the Floquet quasi-energies $|\epsilon_2-\epsilon_1|$ for the dimensionless Hamiltoninan given by Eq.~(\ref{dimlessH}) as a function of $A$ and $\Delta.$  b) The LZS parameter $P_{LZS}=1-\exp (-\pi\Delta ^{2}/2A)$ as function of $A$ and $\Delta.$}
\label{Floqfig}
\end{figure}
%%%%%%%%%%%%%%%%%%%%%%%%%%%%%%%%%%%%%%%%%%%%%%%%%%%%%%%%%%%%%%%%%%%%%%%%%%%%

Depending on the parameters, $H_s$ can induce diverse dynamical behaviors, which are qualitatively different. In order to reduce the number of the parameters, we introduce a dimensionless time variable $\tau=\Omega t$, and obtain that without environmental effects, the Schr\"{o}dinger equation
\begin{equation}
i \frac {d}{d \tau}|\Psi\rangle=\left(A \cos(\tau)\sigma_z +\frac{\Delta}{2}\sigma_x\right) |\Psi\rangle
\label{dimlessH}
\end{equation}
governs the dynamics, with $A=a/\Omega$ and $\Delta=\delta/\Omega$. To compare with the LZS model, the parameter that characterizes the dynamics can be $P_{LZS}=1-\exp (-\pi\Delta ^{2}/2A)$, which would be the transition probability if the external field were linear with a sweep rate equal to the maximal one (at the crossing), i.e., $A$. From a different point of view, strong and weak driving also results in qualitatively different dynamical behavior, here $A/\Delta$ is the relevant parameter: when $A$ is small compared to $\Delta$, the populations in the eigenbasis of $\sigma_x$ exhibits oscillations with the (dimensionless) Rabi frequency $\Omega_R=\sqrt{A^2+(1-\Delta)^2}$, while a strong driving field $A\gg\Delta$ forces $2 \pi$-periodic dynamics. Additionally, when $A<\Delta,$ it is clear from the optical analogy, that RWA in the system-field interaction accurately describes the dynamics. Note that in our case this approximation (dropping fast oscillating counter-rotating terms in $\widetilde{H_s}$, i.e., replacing terms proportional to $\cos \tau$ in the offdiagonals with $1/2 \exp \pm i\tau$) is equivalent to assuming a rotating external field \cite{CCG05} in the $y-z$ plane:
\begin{equation}
H_{s}^{\mathrm{RWA}}=A/2 \cos(\tau)\sigma_z +A/2 \sin(\tau)\sigma_y.
\label{HRWA}
\end{equation}

Fig.~\ref{Floqfig} shows the difference of the Floquet quasi-energies $|\epsilon_2-\epsilon_1|$ for the dimensionless Hamiltonian appearing in Eq.~(\ref{dimlessH}) as a function of $A$ and $\Delta.$ For $H_{s}^{\mathrm{RWA}},$ the difference $\epsilon_1-\epsilon_2$ can be calculated analytically and, clearly, it is equal to the Rabi frequency $\Omega_R.$ As expected, RWA is an accurate approximation for $A\ll\Delta$: Around driving field amplitudes $A=\Delta,$ the relative difference between the quasi-energies with and without RWA is of the order of $10\%,$ and it is increasing for larger values of $A.$ Note that parameters when $\epsilon_2-\epsilon_1=n, n=0,1,\ldots$ are special in the sense %%%%%%%%%%%%%%%%%%%%%%%%%%%%%%%%%%%%%%%%%%%%%%%%%%%%%%%%%%%%%%%%%%%%%%%%%%%%
\begin{figure}[tbp]
\begin{center}
\includegraphics*[width=8.5 cm]{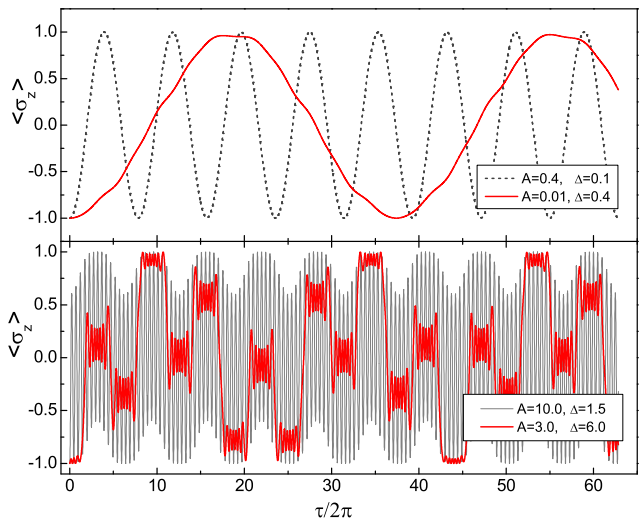}
\end{center}
\caption{Decoherence-free time evolution of the expectation value $\langle \sigma_z\rangle$ (without decoherence) for different parameter values. The initial state is the lower eigenstate of $\sigma_z$, $\rho_s(0)=|-\rangle\langle-|.$}
\label{cohevol}
\end{figure}
%%%%%%%%%%%%%%%%%%%%%%%%%%%%%%%%%%%%%%%%%%%%%%%%%%%%%%%%%%%%%%%%%%%%%%%%%%%%
that in these cases the dynamics has the same periodicity as the driving field. Fig.~\ref{Floqfig} also shows the LZS parameter $P_{LZS},$ indicating the relation between the LZS-type classification of the parameter space and the cases of strong and weak driving. Figure \ref{cohevol} shows the time evolution of $\langle\sigma_z\rangle$ for four qualitatively different situations, strong and weak driving with both $A$ and $\Delta$ being either larger or smaller than the driving frequency $\Omega=1.$ The different periodicity and the role of small and large LZS parameter can clearly be seen in these plots.

\subsection{Quasistationary solutions with decoherence}
\label{couplings}
In this section we investigate the steady state solutions of the master equation (\ref{master}), i.e., the case when the time derivatives of the interaction picture matrix elements are zero. In fact, these solutions do depend on time, the matrix elements will not be constants in the Schr\"{o}dinger picture, but this kind of time dependence is well known, and, additionally, considering a Hamiltonian with explicit time dependence, it is clear that generally there are no constant solutions. In fact, steady state solutions of Eq.~(\ref{master}) in the sense above are periodic \cite{BHP00} due to the time evolution of the Floquet states $|u_r(t)\rangle.$

Now we specify the terms $\mathcal{D}(\omega)$ and $g(\omega)$ in $\gamma$ and $\gamma'$ (Eq.~(\ref{gammas})) that describe the mode density of the reservoir and the strength of the coupling of the spin system to the environmental mode with frequency $\omega.$ Assuming a thermal bath, the average number of excitations $\langle n(\omega) \rangle$ is given by  the Bose-Einstein distribution. Recalling the case of a two-level atom in thermal electromagnetic field \cite{WM94} and a molecular nanomagnet in phonon bath \cite{LL00}, the choice
\begin{eqnarray}
\gamma(\omega)&=&\kappa\frac{\omega^3 e^\frac{\omega}{k T_r}}{e^\frac{\omega}{k T_r}-1} , \\ \gamma^{\prime}(\omega)&=&\kappa\frac{\omega^3}{e^\frac{\omega}{k T_r}-1} \label{gammas2},
\end{eqnarray}
is rather general (recall that $\hbar=1$). Here the $\omega$ independent $\kappa$ describes the overall strength of the coupling, and $T_r$ is the temperature of the reservoir. Additionally, it is instructive to rearrange the terms in the master equation (\ref{master}) by collecting the coefficients of the interaction picture matrix elements $\dot{\rho}_{ij}=\langle u_i (0)| d \rho_s / d t |u_j(0)\rangle.$ Focusing on the change of the populations, we obtain a Pauli-type equation
\begin{eqnarray}
\dot{\rho}_{11}&=&\kappa\left[ \rho_{22}\left(\Gamma_{21}+\Gamma'_{12}\right)-\rho_{11}\left(\Gamma_{12}+\Gamma'_{21}\right)\right],
\nonumber \\
\dot{\rho}_{22}&=&-\dot{\rho}_{11},
\label{pauli}
\end{eqnarray}
while the offdiagonal elements decay according to
\begin{eqnarray}
\dot{\rho}_{12}=\dot{\rho}^{*}_{21}=-\frac{\kappa}{2} \left( \Gamma_{11}+\Gamma_{22}+\Gamma_{12}+\Gamma_{21} - 2\Gamma_{3} \right. \nonumber\\
\left.   + \Gamma'_{11}+\Gamma'_{22}+\Gamma'_{12}+\Gamma'_{21} - 2\Gamma'_{3} \right) \rho_{12}.
\label{offd}
\end{eqnarray}

The coefficients on the rhs.~of Eqs.~(\ref{pauli}) and (\ref{offd}) are determined by the type of the system-environment coupling $\mathcal{S}$ and parameters $A$ and $\Delta$ in the following way:
\begin{eqnarray}
\Gamma_{ii}&=&\sum_{n>0} \gamma(\omega^0_{n}) \left|\langle \langle i|S|i\rangle\rangle_n\right|^2, \nonumber \\
\Gamma_{12}&=&\sum_{n\geq 0} \gamma(\omega^+_{n}) \left|\langle \langle 2|S|1\rangle\rangle_{n}\right|^2, \nonumber \\
\Gamma_{21}&=&\sum_{n> 0} \gamma(\omega^-_{n}) \left|\langle \langle 1|S|2\rangle\rangle_{n}\right|^2,\nonumber \\
\Gamma_{3}&=&\sum_{n> 0} \gamma(\omega^0_{n}) \langle \langle 2|S|2\rangle\rangle_{n} \langle \langle 1|S|1\rangle\rangle_{-n},
\label{Gam}
\end{eqnarray}
and the primed quantities (that vanish at zero temperature) can be obtained by the substitution $\gamma \rightarrow \gamma'.$
Setting the lhs.~of Eqs.~(\ref{pauli}, \ref{offd}) to zero, the solution is a diagonal density matrix (in the $|u_1(0)\rangle, |u_2(0)\rangle$ basis), where the ratio of the populations is given by
\begin{equation}
\frac{\rho^{qs}_{11}}{\rho^{qs}_{22}}=\frac{\Gamma_{21}+\Gamma'_{12}}{\Gamma_{12}+\Gamma'_{21}}.
\label{steadystate}
\end{equation}
Clearly, $\rho^{qs}_{11}+\rho^{qs}_{22}=1,$ and the time evolution converges to the solution above in the long time limit: $\rho^{qs}_{ii}=\rho_{ii}(\infty).$ Let us recall that it is possible to interpret $\Gamma_{12}+\Gamma'_{21}$ ($\Gamma_{21}+\Gamma'_{12}$) as the sum of the environment induced $|u_1\rangle \rightarrow |u_2\rangle$ ($|u_2\rangle \rightarrow |u_1\rangle$) transition rates over different number of excitations in the driving field. This implies that for strong driving, the steady state solution at zero temperature will not necessarily be the ground state $|u_1\rangle:$ the coupled system of the spin and the driving field can emit excitations into the reservoir by both transitions $|u_1\rangle \rightarrow |u_2\rangle$, $|u_2\rangle \rightarrow |u_1\rangle$, provided the net energy flow is directed towards the environment. This process is possible in the $|u_1\rangle \rightarrow |u_2\rangle$ (from ground to excited state) transition as well, since the energy gain of the spin system can be compensated by an appropriate loss in the energy of the driving field.

%%%%%%%%%%%%%%%%%%%%%%%%%%%%%%%%%%%%%%%%%%%%%%%%%%%%%%%%%%%%%%%%%%%%%%%%%%%%
\begin{figure}[tbp]
\begin{center}
\includegraphics*[width=8.5 cm]{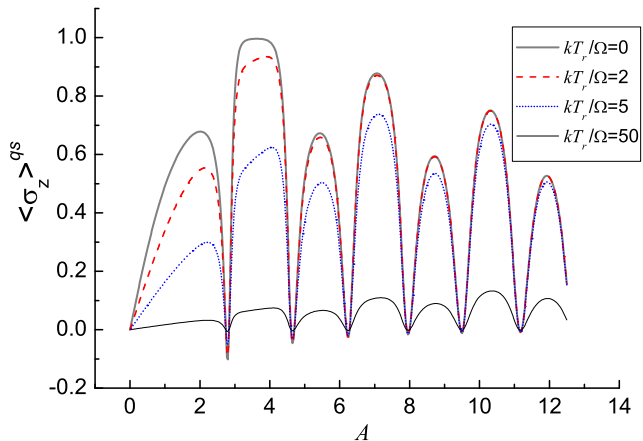}
\end{center}
\caption{The quasistationary expectation value $\langle \sigma_z \rangle^{qs}=\mathrm{Tr}_s\left(\rho^{qs} \sigma_z\right)$ versus the driving field amplitude for $\Delta=1.5$ and different temperatures.}
\label{final}
\end{figure}
%%%%%%%%%%%%%%%%%%%%%%%%%%%%%%%%%%%%%%%%%%%%%%%%%%%%%%%%%%%%%%%%%%%%%%%%%%%%

This effect combined with the parameter dependence of the quasistationary states can be used for {\em state preparation}: As an example, Fig.~\ref{final} shows the quasistationary expectation value $\langle \sigma_z \rangle^{qs}=\mathrm{Tr}_s\left(\rho^{qs} \sigma_z\right)$ as a function of the driving field amplitude for different temperatures. As we can see, there are several points, where the long time limit solution is basically a certain eigenstate of $\sigma_z$ at time instants $\tau=2n \pi$. (For population transfer at periodic crossings using a different method see Ref.~\cite{GS92}.) Note that the validity of RWA in the system-environment interaction (this assumption led us to the master equation (\ref{master}), or, in other words, it resulted in the fact that there is only a single $\omega$ in each term of the sums defining the coefficients $\Gamma$ in Eq.~(\ref{Gam})) requires $|\epsilon_r-\epsilon_{r'}|\gg \gamma(\omega_{n}) \left|\langle \langle i|S|j\rangle\rangle_n\right|^2.$ Clearly, when $\epsilon_1=\epsilon_2,$ this condition cannot be met, but the parameter values where the quasistationary sates are the eigenstates of $\sigma_z$ are far from these degenerate points.

On the other hand, for weak driving, only the ground state $|u_1\rangle$ will be populated in the long time limit. In this case, apart from resonance $\Delta=1/2$, $|u_1\rangle$ is basically the equal weight antisymmetric superposition of the $\sigma_z$ eigenstates, $|u_1\rangle(0)\approx (|+\rangle - |-\rangle)/\sqrt {2},$ implying that in the quasistationary case $\langle\sigma_z \rangle\approx 0.$

For high temperatures $\Gamma_{12} \approx \Gamma_{21},$ thus, as it is expected, in this case the reduced density operator of the spin system will always be proportional to unity in the long time limit. Let us note, however, that temperatures in the mK range can already be termed as high in the context above: if $\Omega$ has the order of magnitude of MHz, $kT_r/\Omega=100$ is satisfied with $T_r\approx 1$ mK.

\subsection{Dynamics: Decoherence time and pointer states}
The results of the previous subsection regarding the quasistationary solutions do not depend qualitatively on the type of the system-environment coupling, but for dynamical calculations, we have to specify  the operator $\mathcal{S}.$ In the present paper we investigate the cases $\mathcal{S}=\sigma_x,\sigma_y,\sigma_z.$

\begin{figure}[tbp]
\begin{center}
\includegraphics*[width=8.5 cm]{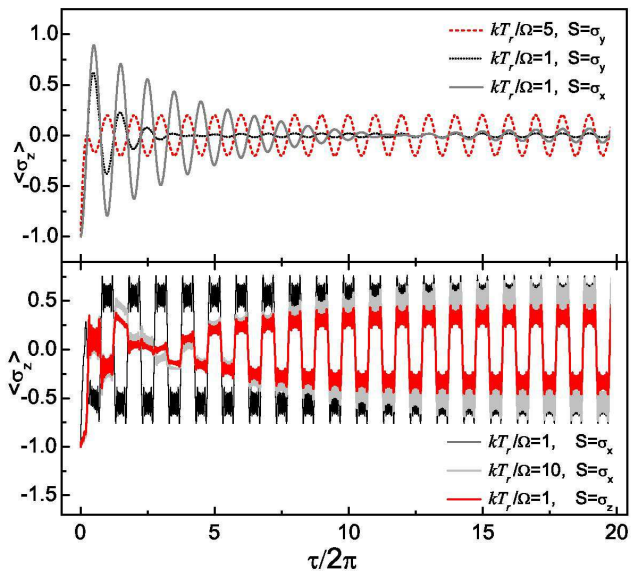}
\end{center}
\caption{Time evolution of the expectation value $\langle \sigma_z\rangle$ for different parameters and decoherence rates ($A=0.1, \Delta=0.5, \kappa=1.0$ for the upper, and $A=10, \Delta=0.4, \kappa=10^{-4}$ for the lower graph), system-environment coupling operators and temperatures. The initial state is the lower eigenstate of $\sigma_z$, $\rho_s(0)=|-\rangle\langle-|.$ Note the emergence of solutions with the same periodicity as that of the driving field.}
\label{decevol}
\end{figure}

Figure \ref{decevol} shows examples of dynamics of the expectation value $\langle\sigma_z \rangle$ for different system-environment coupling strengths, temperatures and a coupling operators. As we can see, the oscillations seen in the free time evolution are damped in this case. The higher the temperature, the stronger this damping effect is. As temperature modifies the final (quasistationary) state according Eq.~(\ref{steadystate}), the amplitude of the long time limit oscillations of $\langle\sigma_z \rangle$ is also temperature dependent.

%%%%%%%%%%%%%%%%%%%%%%%%%%%%%%%%%%%%%%%%%%%%%%%%%%%%%%%%%%%%%%%%%%%%%%%%%%%%
%%%%%%%%%%%%%%%%%%%%%%%%%%%%%%%%%%%%%%%%%%%%%%%%%%%%%%%%%%%%%%%%%%%%%%%%%%%%

In our case Eq.~(\ref{offd}) shows that the final reduced density operator in the relevant interaction picture will be diagonal in the Floquet basis for any type of system-environment coupling. In other words, decoherence drives the system into an incoherent sum of these states, thus they can be considered as pointer states \cite{Z81} in our system. However, there are important differences from the usual picture of decoherence. The first, and probably less notable point is the fact that in our case it is difficult to clearly distinguish between the energy transfer between the investigated system and its environment and decoherence (which, in this context refers to the bare loss of quantum coherence). The time scale of these two, conceptually different processes is roughly the same (mathematically this is reflected by the fact that the rate of change of the diagonal and offdiagonal elements of $\rho_s$ is comparable), thus one can not conclude that first fast decoherence takes place, which is followed by a slow dissipative process leading to thermal equilibrium with the environment. In some sense it is a size effect: in larger systems with more degrees of freedom it is possible to make a dynamical distinction between decoherence and dissipation based on the time scales (see e.g.~Ref.~\cite{FCB01a}). On the other hand, the nature of the problem implies that we have \emph{time dependent pointer states}: even in the long time limit, when $\rho_s$ is diagonal in the Floquet states, $\langle\sigma_z\rangle$ oscillates as a consequence of the time dependence of $|u_1\rangle$ and $|u_2\rangle$.

%%%%%%%%%%%%%%%%%%%%%%%%%%%%%%%%%%%%%%%%%%%%%%%%%%%%%%%%%%%%%%%%%%%%%%%%%%%%
\begin{figure}[tbp]
\begin{center}
\includegraphics*[width=8.5 cm]{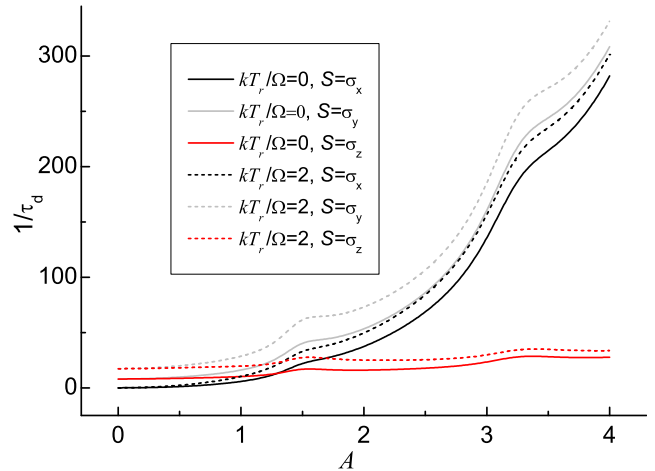}
\end{center}
\caption{The rate of the decoherence, $\tau_d^{-1}$ as a function of the driving field amplitude for different system-environment coupling operators and temperatures.}
\label{timefig}
\end{figure}
%%%%%%%%%%%%%%%%%%%%%%%%%%%%%%%%%%%%%%%%%%%%%%%%%%%%%%%%%%%%%%%%%%%%%%%%%%%%

The characteristic time of the decoherence, $\tau_d,$ according to the considerations above, can be defined as the time instant when the offdiagonal elements of the reduced density operator become smaller then an appropriately chosen percentage of their initial magnitude. Recalling Eq.~(\ref{offd}), this definition can be reformulated as
\begin{eqnarray}
\tau_d=\mathrm{Re} \frac{2}{\kappa} \left( \Gamma_{11}+\Gamma_{22}+\Gamma_{12}+\Gamma_{21} - 2\Gamma_{3} \right. \nonumber\\
\left.   + \Gamma'_{11}+\Gamma'_{22}+\Gamma'_{12}+\Gamma'_{21} - 2\Gamma'_{3} \right)^{-1}.
\label{td}
\end{eqnarray}
Note that the expression above is independent of the initial state of the system, it is valid also in the case when the offdiagonal elements of the density operator are zero at $\tau=0$. Fig.~\ref{timefig} shows $\tau_d^{-1}$ as a function of the driving field amplitude for different system-environment couplings $\mathcal{S}.$ As we can see, the overall tendency is the acceleration of decoherence as the amplitude increases, which is related to the width of the distributions $\langle \langle r'|S|r\rangle\rangle_n,$ i.e., by increasing $A,$ there will be more terms in the coefficients $\Gamma$ that are not negligible. The fine structure of the curves is determined by the amplitude dependence of the Floquet quasi-energies and states, the maxima correspond to the cases of $\epsilon_1\approx \epsilon_2.$  We can also observe that different coupling operators induce different decoherence rates even for the same value of $\kappa.$ Note that this effect can already be seen in Fig.~\ref{decevol}, where the damping of the coherent oscillations were different for $\mathcal{S}=\sigma_x,\sigma_y,\sigma_z.$ It is particularly interesting that for the case of $A$ approaching zero, $S_y$ and $S_z$ lead to a finite decoherence time, while for $S_x$ decoherence effects become negligible in this limit.   This behavior can easily be explained by using the approximate RWA solutions for $|u_1\rangle$ and $|u_2\rangle$ that contain basically two Fourier components, and lead to vanishing (finite) $\Gamma$ coefficients for $S_x$ ($S_y$ and $S_z$).

In an experimental situation, where $\langle\sigma_z \rangle$ can be measured with a high enough temporal resolution, and the coupling of the system to the environment is unknown, the facts discussed above can be used to gain information concerning the nature of this interaction by varying the amplitude and orientation of the external field.

\section{Hysteresis curves: from a series of loops to ladders}

It is often instructive to investigate the response of a physical system subjected to a periodic field as a function of this field itself. Assuming that our two-level system represents a spin 1/2 particle, $\langle\sigma_z \rangle$ is proportional to its magnetization in the $z$ direction, i.e., parallel to the dimensionless driving field $F=A \cos\tau.$  Based on this aspect of the model, and the fact that -- as we shall see -- the functions $\langle\sigma_z \rangle (F)$ are usually multi-valued, these plots can be called hysteresis curves.

%%%%%%%%%%%%%%%%%%%%%%%%%%%%%%%%%%%%%%%%%%%%%%%%%%%%%%%%%%%%%%%%%%%%%%%%%%%%
\begin{figure}[tbp]
\begin{center}
\includegraphics*[width=8.5 cm]{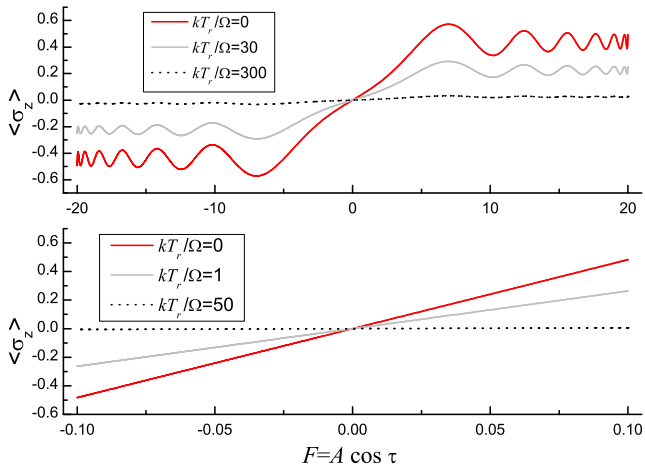}
\end{center}
\caption{Quasistationary magnetization curves for different parameter values ($\Delta=1.5, A=20$ for the upper, $\Delta=0.6, A=0.1$ for the lower curve) and temperatures. Note that for small values of $A,$ the Floquet states have only a few Fourier components that are not negligible, but this number increases for stronger driving.}
\label{stathystfig}
\end{figure}
%%%%%%%%%%%%%%%%%%%%%%%%%%%%%%%%%%%%%%%%%%%%%%%%%%%%%%%%%%%%%%%%%%%%%%%%%%%%

Let us start with the quasistationary solutions of the master equation (\ref{master}), when the density operator is diagonal in the Floquet (pointer) basis, with the populations given by Eq.~(\ref{steadystate}). In this case the periodicity of the Floquet states imply that the hysteresis curves are closed lines. The oscillations of the curves shown in Fig.~\ref{stathystfig} are related to the time evolution of the Floquet states, the more Fourier components these pointer states have, the more local maxima and minima can be observed in the graph $\langle\sigma_z \rangle (F).$

Since for high temperatures the quasistationary solution is proportional to the unit matrix (implying $\langle\sigma_z \rangle = 0$), we expect the functions $\langle\sigma_z \rangle(F)$ to be squeezed in the vertical direction as the temperature increases.  This effect can clearly be seen in Fig.~\ref{stathystfig}, for high enough temperatures the plotted curves are close to a horizontal line, independently from the parameters.

%%%%%%%%%%%%%%%%%%%%%%%%%%%%%%%%%%%%%%%%%%%%%%%%%%%%%%%%%%%%%%%%%%%%%%%%%%%%
\begin{figure}[tbp]
\begin{center}
\includegraphics*[width=8.7 cm]{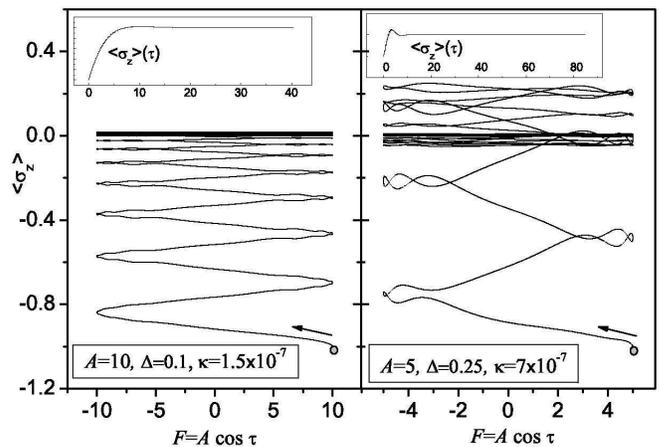}
\end{center}
\caption{Convergence towards the quasistationary magnetization curves at high temperatures ($kT_r/\Omega=600$). The grey dot and the arrow indicate the starting point (corresponding to $\rho_s(0)=|-\rangle\langle-|$) and the initial direction. The right panel shows the case of weak decoherence, while environment induced effects are moderate for the left panel. (Note that according to Fig.~\ref{timefig}, the same value of $\kappa$ means faster decoherence for larger driving field amplitudes.) The insets show $\langle \sigma_z \rangle$ as a function of $\tau /2 \pi.$}
\label{dynhystfighT}
\end{figure}
%%%%%%%%%%%%%%%%%%%%%%%%%%%%%%%%%%%%%%%%%%%%%%%%%%%%%%%%%%%%%%%%%%%%%%%%%%%%

\begin{figure}[tbp]
\begin{center}
\includegraphics*[width=8.5 cm]{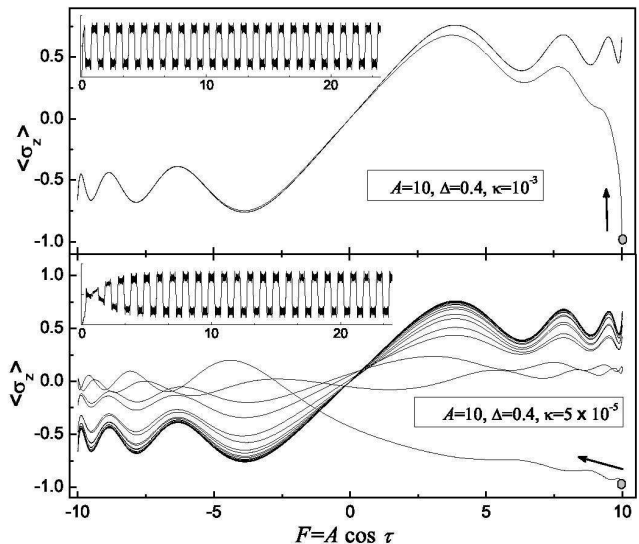}
\end{center}
\caption{Convergence towards the quasistationary magnetization curves at low temperatures ($kT_r/\Omega=1$). The grey dot and the arrow indicate the starting point (corresponding to $\rho_s(0)=|-\rangle\langle-|$) and the initial direction. The insets show $\langle \sigma_z \rangle$ as a function of $\tau/2 \pi.$}
\label{dynhystfiglT}
\end{figure}

Unless the initial density operator of the spin system is equal to the quasistationary solution,  the function $\langle\sigma_z \rangle \left(F(\tau)\right)$ obtained during the whole time evolution $\tau=0 \ldots \infty$ is multi-valued, and this hysteresis curve reflects the convergence of the system to the long time limit solution. That is, as Figs.~\ref{dynhystfighT} and \ref{dynhystfiglT} show, a quasistationary magnetization curve acts as an attractor, all the paths $\langle\sigma_z \rangle \left(F(\tau)\right)$ converge to this curve independently from the initial point $\langle\sigma_z \rangle \left(F(0)\right).$ Visually, the length of the path till the quasistationary curve is reached is related to the dynamics: if there are a lot of detours before getting close to the final curve, decoherence is slow; on the other hand, when only a single line is visible towards the quasistationary curve, coherent oscillations are damped strongly.

Let us note, that in certain experimental situations the dimensionless amplitude $A$ can fall orders of magnitude beyond the applicability of the method described so far. The main problem is the determination of the Floquet eigenstates, which requires the diagonalization of a matrix, the dimension of which is proportional to $A.$ Therefore we developed an alternative, approximate way of calculating the parameters appearing in the dynamical equations (\ref{pauli}, \ref{offd}): The eigenvalue equation that leads to the Fourier components
\begin{equation}
\langle \pm|u_i\rangle_n={\int}_0^{T} e^{(i n \Omega t)}\langle \pm|u_i(t)\rangle_n dt
\end{equation}
can be rewritten as a system of four differential equations with $n$ being considered as a continuous variable. These equations are coupled via a term which is proportional to $\Delta/A,$ which can usually be considered as a small number, allowing for the equations to be solved iteratively. The zero order approximation (assuming $\Delta/A=0$) is given in terms of the Airy functions
\begin{eqnarray}
\label{eigapprox}
\langle +|u_1\rangle_n&=&N_+ \mathrm{Ai}\left(-\left(\frac{2}{A} \right)^\frac{1}{3} (n+A)\right), \nonumber \\
\langle -|u_1\rangle_n&=&N_- \mathrm{Ai}\left(\left(\frac{2}{A} \right)^\frac{1}{3} (n-A)\right), \nonumber \\
\langle +|u_2\rangle_n&=&\langle -|u_1\rangle_n, \ \ \langle -|u_2\rangle_n=-\langle +|u_1\rangle_n,
\end{eqnarray}
where $N_\pm$ are constants. The next iteration provides solutions where the oscillatory part of the Airy functions (for negative arguments) are also strongly damped for $|n|>A.$ Using these analytical approximations, we found that in the high temperature limit, the coefficients $\Gamma$ are proportional to $A^2$ (indications to this kind of behavior can already be seen in Fig.~\ref{timefig}). Note that these analytical results were verified by numerical calculations with the largest amplitudes allowed by our computational resources. In fact, not only the scaling with $A^2$ were seen, but a good agreement concerning the prefactor has also been found. Similarly, the approximate eigenstates (\ref{eigapprox}) in those domains where the Airy functions do not oscillate too fast were also found to be in a reasonable agreement with the numerically exact results.

%%%%%%%%%%%%%%%%%%%%%%%%%%%%%%%%%%%%%%%%%%%%%%%%%%%%%%%%%%%%%%%%%%%%%%%%%%%%
%%%%%%%%%%%%%%%%%%%%%%%%%%%%%%%%%%%%%%%%%%%%%%%%%%%%%%

\subsection*{Application to molecular nanomagnets in slowly oscillating, large amplitude external magnetic fields}
At high temperatures, when $\langle\sigma_z \rangle=0$ is the final solution, the hysteresis curves can reach the final horizontal line via a series of steps (see the left panel of Fig.~\ref{dynhystfighT}). According to a recent experimental result \cite{M03}, this kind of behavior can be relevant in physical systems where crystals consisting of high-spin molecules such as Mn$_{12}$-Ac (or simply Mn$_{12}$) and Fe$_{8}$ (also known as molecular nanomagnets \cite{GSV06}) are being driven by periodic external magnetic fields. These special molecules contain transition metal atoms with strongly exchange-coupled spins, which causes them to behave as a single, large spin. Experiments on the magnetization dynamics of these molecular crystals have shown the presence of a series of
steps in the magnetization curve at sufficiently low temperatures \cite{TL96,FST96,MSS01,WMC06}. This behavior is a consequence of quantum mechanical tunneling of spin states through the anisotropy energy barrier and occurs when the external field brings two levels at different sides of the barrier into resonance via Zeeman interaction. When the external magnetic field is swept linearly, an appropriate LZS model around a certain resonance provides a very useful approximate description of the dynamics (physical consequences of the difference between a model involving not only two energy levels and the LZS treatment can be found in Ref.~\cite{PP07}).

In the following we focus on the molecule Mn$_{12}$, which can be
considered as a representative example of molecular nanomagnets, and, consequently, it has been investigated in several important experimental works, including the one reported in Ref.~\cite{M03}. In this experiment the sweep rate was 5.83 mT/s, with amplitudes around 0.25 T, while the temperature was $T_r=0.25$ K, corresponding to the high temperature limit $kT_r/\hbar\Omega\gg 1.$ As an example, let us concentrate on the seventh resonance around 3.67 T, where the magnetic levels labeled by $m=-10$ and $m'=3$ correspond to the $\sigma_z$ eigenstates $|-\rangle$ and  $|+\rangle$ in our model, respectively. The level splitting at this anticrossing can be calculated using the appropriate spin Hamiltonian \cite{LL00,PP07}, leading to $7 \times 10^{-7}$ K in temperature units. Combining these values, the dimensionless parameters are $A\approx 10^{12},$ $\Delta\approx 10^{6},$ thus the approximate method described previously in this section has to be applied.

%%%%%%%%%%%%%%%%%%%%%%%%%%%%%%%%%%%%%%%%%%%%%%%%%%%%%%%%%%%%%%%%%%%%%%%%%%%%%%%%%%%%%%%%%%%%%%%%%%%%%%%%%
\begin{figure}[tbp]
\begin{center}
\includegraphics*[width=8.5 cm]{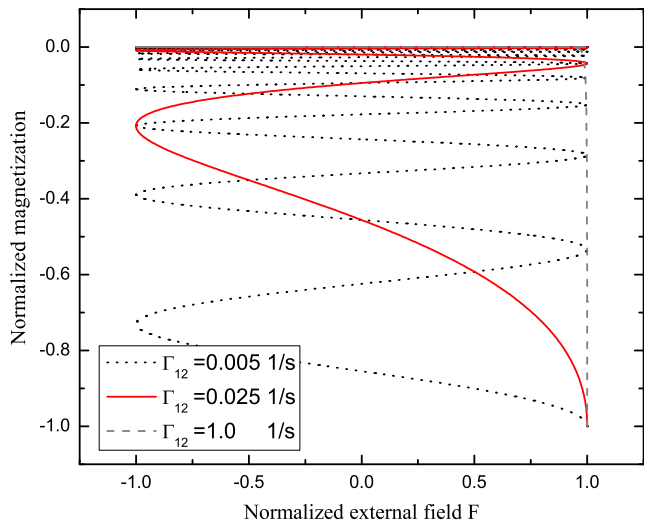}
\end{center}
\caption{Hysteresis curves for slowly oscillating external fields calculated with assuming incoherent transitions as a consequence of dephasing. The figure corresponds to the seventh resonance (around 3.67 T) of the molecular nanomagnet Mn$_{12}$ driven by an external field with amplitude of 0.25 T and $\Omega=0.1$ 1/s. The ladder-like curve for the weakest phonon induced effect (dotted line) reflects qualitatively the results of Ref.~\cite{M03}.}
\label{lastfig}
\end{figure}
%%%%%%%%%%%%%%%%%%%%%%%%%%%%%%%%%%%%%%%%%%%%%%%%%%%%%%%%%%%%%%%%%%%%%%%%%%%%%%%%%%%%%%%%%%%%%%%%%%%%%%%%%

Physically, the overall decoherence rate (resulting from various effects \cite{GSV06}) in Mn$_{12}$ is around $10^{6}$--$10^{8}$ 1/s, which means an extremely fast process compared to the oscillation of the external field, where $\Omega\approx 0.1$ 1/s. If we assume that the phonon bath is the only source of decoherence (which is not the case, see below), it is impossible to obtain ladder-like hysteresis curves similar to the experimental results. For very strong phonon induced decoherence, $\langle\sigma_z\rangle$ as a function of the external field would show a sudden convergence towards its stationary value of zero, in other words, the hysteresis curve would consist of two, almost perpendicular straight lines, starting with a vertical one that connects the initial point and the stationary horizontal line. (Similarly to the curve corresponding to the strongest decoherence in Fig.~\ref{lastfig}.) However, both experimental and theoretical results show (see eg.~Refs.~\cite{W07} and \cite{MST06}) that the main source of decoherence is related to dipolar and hyperfine interactions which modify the local environment of the spins and eventually result in a distribution of the level splittings. Investigating the dynamical properties of our system, it can be seen (both analytically and numerically) that this effect decreases the offdiagonal elements of the density matrix in the Floquet basis. The characteristic time of the process depends on the width of the distribution of the level splittings, but if we take into account that a single period of the external field takes roughly 10 s, dephasing due to dipolar and hyperfine interactions is practically instantaneous. We note, however, that this kind of dephasing usually does not lead to a horizontal stationary magnetization curve, as even the incoherent sum of the Floquet states has nontrivial time dependence.

This implies that it is worth assuming that the results reported in Ref.~\cite{M03} reflect the interplay of fast dephasing due to the distribution of the level splittings and a much slower phonon induced process. Therefore we calculated the dynamics with constantly zero offdiagonal density matrix elements in the Floquet basis (corresponding to instantaneous dephasing) by the aid of the approximations that led us to Eqs.~(\ref{eigapprox}). As it is shown by Fig.~\ref{lastfig}, the value of $\Gamma_{12}=0.005$ 1/s leads to qualitative agreement with the experimental results. However, it is clear that decoherence rates $\Gamma$ are not variable parameters, their values are determined by the physical system we are considering. Concretely, the value of $\kappa$ in Eq.~\ref{gammas2} (that can be calculated according to Ref.~\cite{LL00}) and the scaling of $\Gamma(A)$ as $A^2$ together determine the decoherence rates. In this procedure there are basically no free parameters (apart from some uncertainty of the sound velocity in the molecular crystal) and we obtained that $\Gamma_{12}$ is around 0.025 1/s, which, according to Fig.~\ref{lastfig}, is slightly too large, it leads to hysteresis curves with a few steps only. However, keeping in mind the approximate nature of our treatment (considering not only the calculation of the Floquet states (\ref{eigapprox}), but also the spin-phonon coupling operator, see eg.~Ref.~\cite{LL00}), we think that the methods developed in this paper provide a solid starting point for a theory that aims to describe nanomagnets driven by periodic magnetic fields in a quantitative way. Concerning our current results, we can conclude that the thermal phonon bath alone can not be responsible for the experimentally observed ladder-like hysteresis curves, but taking the strong dephasing also into account, this kind of behavior can be explained.

\section{Conclusions}
We investigated a two-level system which is driven by periodic external field and which is also in interaction with a thermal bath. It has been found that -- independently from the decoherence rate, from the initial state and even from the type of the system-environment coupling -- the time evolution is directed towards an appropriate incoherent sum of periodic Floquet states. The final ratio of the populations related to these time dependent pointer states is determined by the parameters of the system Hamiltonian, the type of the system-environment coupling and the temperature. Our results show that the form of the quasistationary hysteresis curves is completely determined by the time evolution of the Floquet states, while the rate of convergence towards these curves is related to the characteristic time of the decoherence. As an important example of the possible applications, we have shown that our model can be used to describe molecular nanomagnets driven by periodic external magnetic fields.

\label{conclusionsec}

\section*{Acknowledgments}
This work was supported by the Flemish-Hungarian Bilateral
Programme, the Flemish Science Foundation (FWO-Vl), the Belgian Science Policy
and the Hungarian Scientific Research Fund (OTKA) under Contracts Nos.~T48888,
M36803, M045596. P.F.~was supported by the J.~Bolyai grant of the
Hungarian Academy of Sciences.

\end{document}